\documentclass[reprint,secnumarabic,amssymb, nobibnotes, aps, prd]{revtex4-2}
\usepackage{graphicx}
\usepackage{amsmath}
\usepackage{epstopdf}

\begin{document}
\title{Electronic Kapitza conductance and related kinetic coefficients at
an interface between n-type semiconductors}
\author{A.\,P. Meilakhs}
\email[A.\,P. Meilakhs: ]{mejlaxs@mail.ioffe.ru}

\affiliation{$^{1}$Ioffe Institute, 26 Politekhnicheskaya, St. Petersburg 194021,
Russian Federation~~\\
 $^{2}$DFMC, Centro Atomico Constituyentes, CNEA, Av. Gral. Paz 1499,
San Martin, Buenos Aires, 1650, Argentina }
\date{\today}

\begin{abstract}
We calculate the Kapitza conductance, which is the proportionality
coefficient between heat flux and temperature jump at the interface,
for the case of two conducting solids separated by the interface.
We show that for conducting solids in a non-equilibrium state, there
should also arise the electrochemical potential jump at the interface.
Hence to describe linear transport at the interface we need three
kinetic coefficients: interfacial analogs of electric and heat conductances
and interfacial analog of the Seebeck coefficient. We calculate these
coefficients for the case of an interface between n-type semiconductors.
We perform calculations in the framework of Boltzmann transport theory.
We have found out that the interfacial analog of the Seebeck coefficient
for some range of parameters of the considered semiconductors, has
a high value of about $10^{-3}$ V/K. Thus this effect has the potential
to be used for the synthesis of effective thermoelectric materials. 
\end{abstract}

\maketitle

\section{Introduction}

When heat flows through an interface between materials a temperature
jump occurs at the interface. A proportionality coefficient between
the heat flux and the temperature jump is called Kapitza conductance
\cite{Kap}. After the discovery of the phenomena, it was very soon
realized that the temperature jump is due to the reflection of phonons
at the interface \cite{Khal}. The easiest model to calculate the
Kapitza conductance is the Diffuse Mismatch Model (DMM) \cite{Swar1}.
It assumes that interfacial scattering is so strong that the phonon
incident on the interface "forgets"\ its initial direction, and
it is scattered uniformly in all directions \cite{Swar2}.

Nowadays the science of Kapitza conductance is developed in many ways.
Some papers are concerned with improving understanding of the dynamics
of the crystal lattice at the interface with computer simulations
\cite{Din1,Din2,Din3,Din4,DinDMM,Din5} or analytically \cite{AnDin1,AnDin2,AnDin3}.
Others study phonon kinetics at the interface \cite{PhonKin1,PhonKin2,PhonKin3,PhonKin4}.
Often the nonequilibrium Greens function method is used for calculations
\cite{PhonGreen1,PhonGreen2,PhonGreen3}. Not only the theory is developed,
but also new experiments are perpetually conducted \cite{Exp1,Exp2,Exp3,Exp4}.
The reason for such a development is not only the intrinsic interest
of any discontinuous phenomena in physics, but also the importance
for applications \cite{Appl1,Appl2,Appl3}.

Nowhere in the Kapitza conductance researches, very specific properties
of phonons are used. The property that gives raise to the temperature
jump at the interface is the reflection of phonons. That is just the
consequence of phonons being waves. Since electrons are waves too,
they also reflect at the interface. This should give rise to a temperature
jump between two electronic subsystems separated by an interface.

Works on electron transport phenomena in the local region mostly originated
from the seminal manuscript by Landauer \cite{Landauer}, which describes
one-dimensional transport in a disordered media. Later, his ideas
were generalized to three dimensions \cite{Electrons1}. Further developments
included taking into account an external magnetic field \cite{Electrons2},
an electron-electron interaction in the interfacial region \cite{Electrons3,Electrons4},
and some improvementson the numerical methods of calculation \cite{Electrons5,Electrons6}.
These approaches are thoroughly reviewed in refs. \cite{Electrons7,Electrons8}.
However, all these papers only investigate electrical current and
not heat flux. They do not use the notion of a sharp jump, which is
very natural in the context of interfacial kinetics. Also, almost
all papers about electronic transport at the interface use Green's
function formalism and the Kubo formula for calculations.

Here we propose a formalism based on Boltzmann kinetic equation, which
was developed for phonon heat transfer through the interface \cite{Me,KinDMM}.
For electrons, not only the energy density and energy flux, but also
the electric charge and electric current are conserved. More conservation
laws cause two jumps at the interface: the temperature jump and the
electrochemical potential jump. So there are four coefficients that
relate currents through the interface to jumps at the interface. By
the Onsager relations, two of them are equal \cite{Zaim}. So we have
three kinetic coefficients: the interfacial analogs of electric and
heat conductances and thermoelectric coefficient. To the author's
best knowledge, the jump of electrochemical potential at the interface
is introduced here for the first time. Two different arguments for the
existence of such a jump would be provided in Sections 2 and 3.

Kinetic coefficients of the interfaces are important in nanostructures,
where homogeneous parts are small and the interfaces are very close
to each other. For example, the kinetics of interfaces are important
for understanding the kinetics of superlattices. To estimate the contribution
of interfaces to the transport properties of the nanostructure, we
divide the interfacial conductance by the conductance of the one of
homogeneous parts of the material. Such quantities with a dimension
of length, are referred to as Kapitza lengths \cite{KLeng1}. Typical
values of Kapitza lengths are of the order of $100$ nm \cite{KLeng2}.
To produce a superlattice, for which interface kinetical coefficients
are important, layers thickness should be about Kapitza lengths or
thinner.

Quantum particles in superlattices that consist of very thin layers,
only several atomic layers wide, belong to a few layers at once, which
modify their characteristics substantially. This modification was
studied for phonons \cite{SupLat1,SupLat2} and electrons \cite{SupLat3}
for a long time, and nowadays some very beautiful results are obtained
\cite{SupLat4}.

The results of this paper are applicable to other type of structure,
where electrons belong to only one layer and can be reflected, with
some probability, from the interface between layers. Layers in such
a superlattice should be sufficiently thick, such that both, typical
electron wavelength and mean free pass should be less than the width
of one layer. Together with previous estimates of Kapitza lengths,
the superlattices in consideration should have a layer thickness in
the range of approximately tens to hundreds of nanometers.

Superlattices with such properties were considered in a series of
papers \cite{SupLat10,SupLat5,SupLat6,SupLat7}. However, in those
works the reflection of electrons at the interfaces between layers
was treated like a modification of an effective number of particles,
that are involved in the transport process. We think that reflection
should be treated as a cause of Kapitza jumps at the interface. This
approach was used in papers \cite{SupLat8,SupLat9}. However, in these
papers, authors only use temperature jumps and not electrochemical
potential jumps. The paper \cite{SupLat8} is the unique one known
to the author, that directly refers to the temperature jump between
electronic subsystems separated by an interface, but no calculation
of Kapitza conductance was provided in that case.

Superlattices are one of the promising candidates for the fabrication
of thermoelectric materials with high figures of merit \cite{SupLat11}.
This is due to the so-called "electron filtering"\, effect \cite{SupLat10},
which only allows the electrons of high energy to go through the interface.
This effect is of great practical importance since during the last
decade thermoelectricity has turned \cite{ThermEl1} into one of the
most important subjects of applied physics \cite{ThermEl2,ThermEl3}.
The goal is to produce a thermoelectric generator \cite{ThermEl4}
with the largest possible figure of merit $ZT$, and some new principles
were developed to enhance the value of this parameter \cite{ThermEl5}.
Recent advances and more literature can be found in the review \cite{ThermEl6}.

In this paper, we calculate coefficients characterizing transport
through the interface by electrons for the interface between n-type
semiconductors. We will show how the interfacial thermoelectric effect
can be used for the production of a material with very high values
of thermoelectric coefficient.

\section{General theory}

In the kinetic theory of homogeneous media, we have this type of relation
between gradients and flows (the notation is taken from the book \cite{Zaim}):
\begin{align}
q=L_{TT}\nabla T+L_{TE}\nabla U^{*}\nonumber \\
j=L_{ET}\nabla T+L_{EE}\nabla U^{*}.\label{EqLs}
\end{align}
Here $U^{*}$ is the effective electric potential, that is, electrochemical
potential divided by the electron charge $U^{*}=\mu/e+U=\zeta/e$.
With all four $L$-s we can express all measurable kinetic coefficients,
such as conductivity, thermal conductance, Seebeck, and Peltier coefficients
\cite{LdKin}.

To describe interfacial kinetic phenomena, we can write down analogous
equations. We switch from gradients in Eq. \ref{EqLs} ($\nabla$-s)
to finite differences ($\Delta$-s) in the following equation. Also
we write inverse proportionality coefficients, since those are the
ones that can be calculated naturally, as we will see in the next
section. 
\begin{align}
\Delta T=M_{TT}\,q+M_{TE}\,j\nonumber \\
\Delta U^{*}=M_{ET}\,q+M_{EE}\,j.\label{EqMs}
\end{align}

Based on non-equilibrium thermodynamics, we provide here the first
argument for the existence of the electrochemical potential jump.
The electric current and the heat current are thermodynamical flows.
The temperature jump at the interface is a thermodynamic force. We
have a proportionality coefficient between the temperature jump and
the electronic flux ($M_{TE}$). Because of Onsager relations, we
should have a dual coefficient, that relates heat flux with some electronic
force ($M_{ET}$). This force is an electrochemical potential jump
$\Delta U^{*}$. Because of the Onsager reciprocal relations we also
get $M_{TE}=TM_{TE}$.

Here we note one more difference between the homogeneous and the interfacial
cases. In homogeneous media, in the stationary case, $\nabla U^{*}$
is actually just an ordinary electric field $E$: the current flows
through an electro-neutral media, and charge can not be stored anywhere.
In the case of an ideally sharp interface, $\Delta U^{*}$ is actually
$\Delta\mu/e$, i.e. a jump of chemical potential. This is becausea
finite jump of electric potential on a zero distance means an infinite
electric field. For the non-ideal interfaces, of a finite length,
there can be contributions of both $\Delta\mu/e$ and $\Delta U$.
We will not take into account the difference between $\mu/e$ and
$U$, since from the point of view of kinetics they are indistinguishable
\cite{Zaim}.

We should also note here that we only perform the calculations of
linear responses. The interface between semiconductors is known to
generate nonlinear current-voltage characteristics \cite{Anselm}.
The typical expression is for the current is $I\sim\exp(eU/kT)-1$,
where $U$ is the voltage applied to the interfacial layer. Here we
only describe the region of parameters $kT\gg eV$, so the current-voltage
characteristic is linear: $I\sim U$. In our notation, this means
\begin{align}
 & \Delta T\ll T\nonumber \\
 & \Delta U^{*}\ll kT/e.\label{CondLin}
\end{align}

\begin{figure}[t]
\centering \includegraphics[width=0.5\textwidth]{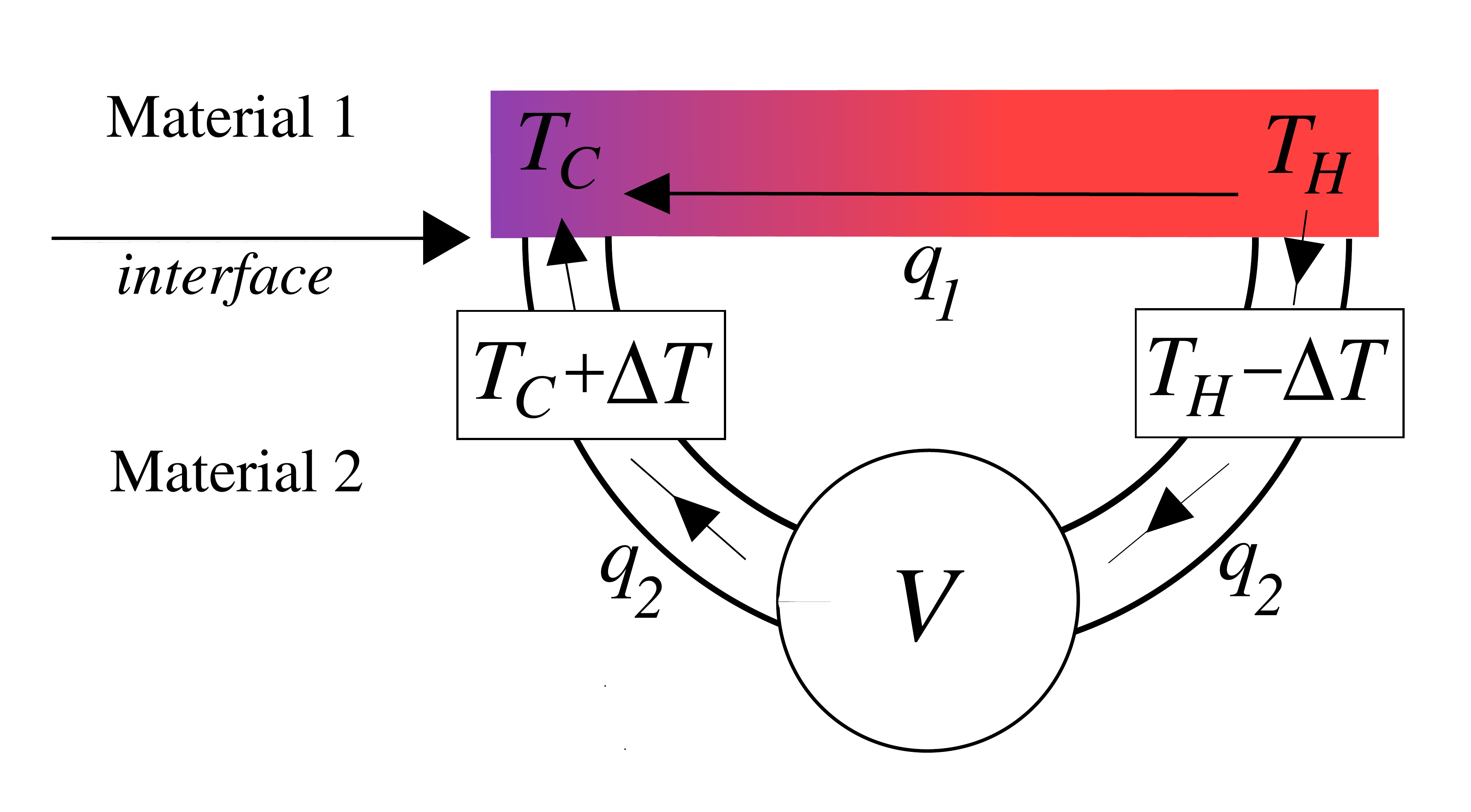} \caption{Sketch of the system under measurement (material 1): one of its ends
is a heat source and the other one is a heat drain. The source is
kept at temperature $T_{H}$ and the drain at temperature $T_{C}$.
The heat flux $q_{1}$ goes through material 1. The measuring device
is shown in white (material 2). The heat flux through the device is
denoted as $q_{2}$. This flux goes through the interface between
materials 1 and 2. Because of the temperature jumps at the interface,
the boundaries of material 2 have temperatures $T_{H}-\Delta T$ and
$T_{C}+\Delta T$, respectively.}
\label{Fig1} 
\end{figure}

We introduced parameters, $M$-s, in formulae (\ref{EqMs}) so that
they would be the most natural quantities to calculate theoretically.
In the next section, we will present how to perform the calculation
{[}Revised until here{]}. Now we want to introduce parameters that
would be the most convenient for experimental measurements. In analogy
with the homogeneous case, we want to introduce three such parameters.
The thermal Kapitza conductance $K_{T}$ is analogous to heat conductance
and is measured under condition $j=0$. The electrical Kapitza conductance
$K_{E}$ is analogous to electric conductivity and is measured under
condition $\Delta T=0$. Finally, the analog of Seebeck coefficient
$K_{S}$, which is proportionality between $\Delta T$ and $\Delta U^{*}$,
under the assumption that $j=0$. Substituting mentioned conditions
into (\ref{EqMs}) yields 
\begin{align}
 & K_{T}=1/M_{TT}\nonumber \\
 & K_{E}=\left(M_{EE}-M_{ET}*M_{TE}/M_{TT}\right)^{-1}\nonumber \\
 & K_{S}=M_{ET}/M_{TT}.\label{EqKs}
\end{align}
Those are desired measurable coefficients. Let us consider the third
one since its use has some interesting issues.

Thermoelectricity, i.e. Seebeck and Peltier effects, is often referred to
as a contact phenomenon. In the case of the Peltier effect, heat fluxes
are generated by electric fields inside the media, but it is the interfacial region
 between the materials that is actually being heated. In the case of Seebeck's effect, it
can only be measured with respect to some other material.
This is so because the material of the measuring device experiences
just the same temperature difference between its ends,
as the material that is being measured. We want to stress the difference between the phenomenon 
of temperature and chemo-electric potential jumps at the interface
from these aspects of the usual Peltier and Seebeck effects. To do so we want
to consider the classic thermocouple experiment, as it is presented in textbooks, such as \cite{Zaim}, and take into account interfacial temperature and chemo-electric potential jumps.

The setup is presented in Fig. \ref{Fig1}. Let's consider a material that is being measured, we call it material 1. The hot end is kept at temperature $T_{H}$ and the cold one is set to temperature $T_{C}$. The piece of material is electrically isolated, so there is no electric current in it ($j=0$).
Since the system is static, but not in equilibrium, there is a constant heat current $q_{1}$. Now we attach the measuring device made of another material, material 2. We
denote $l_{1},\,l_{2}$ the lengths and $\kappa_{1},\kappa_{2}$ the
heat conductivities of materials 1 and 2, respectively. Thermal resistances of these bulk materials are expressed as $l_{1}/\kappa_{1}$ and $l_{2}/\kappa_{2}$.

In general, the thermal resistance of material 2 is different from
the thermal resistance of material 1. So under the same temperature
difference, heat flux through this device is going to be different.
We denote the flux through the measuring device as $q_{2}$. Since
the ends of material 1 are kept at a constant temperature, we can
think about them as the heat source and heat sink. Since the source
and the sink are placed in material 1 and because of the continuity
of heat flux outside of the sink and the source, the heat flux through
the interfaces between the materials will be the same as through material
2, that is $q_{2}$. Considering the temperature and electrochemical
potential jumps at the interface, the voltage measured by device $V$
is 
\begin{equation}
V=2M_{ET}q_{2}-\int\limits _{T_{C}+\Delta T}^{T_{H}-\Delta T}\alpha_{2}\,dT+\int\limits _{T_{C}}^{T_{H}}\alpha_{1}\,dT.\label{EqCoupleStart}
\end{equation}
$M_{ET}$ is the interfacial thermoelectric coefficient from Eq. (\ref{EqMs}),
the first integral goes through the temperature difference between
the ends of material 2 (Fig. \ref{Fig1}), and the second through
material 1.

We neglect temperature dependance of $\alpha_{1},\alpha_{2}$ so we
can rewrite the previous equation 
\begin{equation}
V=2M_{ET}q_{2}+2\Delta T\alpha_{2}+\int\limits _{T_{C}}^{T_{H}}(\alpha_{1}-\alpha_{2})\,dT.
\end{equation}
Since $\Delta T$ is proportional to q we can further rewrite it,
with help of Eq. (\ref{EqMs}): 
\begin{equation}
V=2(M_{ET}+M_{TT}\alpha_{2})q_{2}+(T_{H}-T_{C})(\alpha_{1}-\alpha_{2}).\label{EqCouple_q}
\end{equation}
The second term on the right side is the well-known Seebeck effect.
The first term is the additional contribution from interfaces.

Now we want to express the heat flux in terms of temperature. So we
write 
\begin{equation}
T_{H}-T_{C}=2M_{TT}q_{2}+l_{2}\kappa_{2}^{-1}q_{2}.
\end{equation}
Here we find $q_{2}$ and substitute it into Eq. (\ref{EqCouple_q}),
and finally, we arrive to 
\begin{equation}
V=\left(\frac{K_{S}+\alpha_{2}}{1+\frac{2K_{T}}{l_{2}^{-1}\kappa_{2}}}+\alpha_{1}-\alpha_{2}\right)(T_{H}-T_{C}).\label{EqCoupleFin}
\end{equation}
Here we have used expressions (\ref{EqKs}). We can see that experimentally
measurable quantity $V$ is indeed expressed with $K$-s.

The big fraction in expression (\ref{EqCoupleFin}) represents the
contribution of boundaries to the total measured voltage. Looking
at its denominator we can clearly see, that if the heat resistivity
of the boundaries is much smaller than the resistivity of the inner
part of the second material, this contribution is negligible. In the
opposite case, the contribution to the thermoelectric effect of the
second material is replaced by the contribution of boundaries 
\begin{equation}
V=\left(K_{S}+\alpha_{1}\right)(T_{H}-T_{C}).
\end{equation}

If the heat source and heat drain are placed on the other sides of
the interfaces, inside the material $2$, we will have a different
expression. However, we can obtain it from (\ref{EqCoupleFin}) just
by changing indexes and signs for every kinetic coefficient. We can
have a more interesting result if the heat source is placed in material
1 and the heat drain is placed in material 2 (or otherwise). For this
case, we can rewrite the expression (\ref{EqCoupleStart}): 
\begin{equation}
V=M_{ET}q_{2}-M_{ET}q_{1}-\int\limits _{T_{C}}^{T_{H}-\Delta T}\alpha_{2}\,dT+\int\limits _{T_{C}+\Delta T}^{T_{H}}\alpha_{1}\,dT.
\end{equation}
With the same operations as previously, we obtain 
\begin{equation}
V=\left(\frac{K_{S}+\alpha_{2}}{1+\frac{K_{T}}{l_{2}^{-1}\kappa_{2}}}-\frac{K_{S}+\alpha_{1}}{1+\frac{K_{T}}{l_{1}^{-1}\kappa_{1}}}+\alpha_{1}-\alpha_{2}\right)(T_{H}-T_{C}).
\end{equation}
Again, if the lengths of materials are sufficiently large, we will
have just a classic formula for thermocouples. However, if the lengths
are small, we will get: 
\begin{equation}
V=(K_{S}-K_{S})(T_{H}-T_{C})=0.
\end{equation}

\section{Calculations}

\begin{figure}[t]
\centering \includegraphics[width=0.4\textwidth]{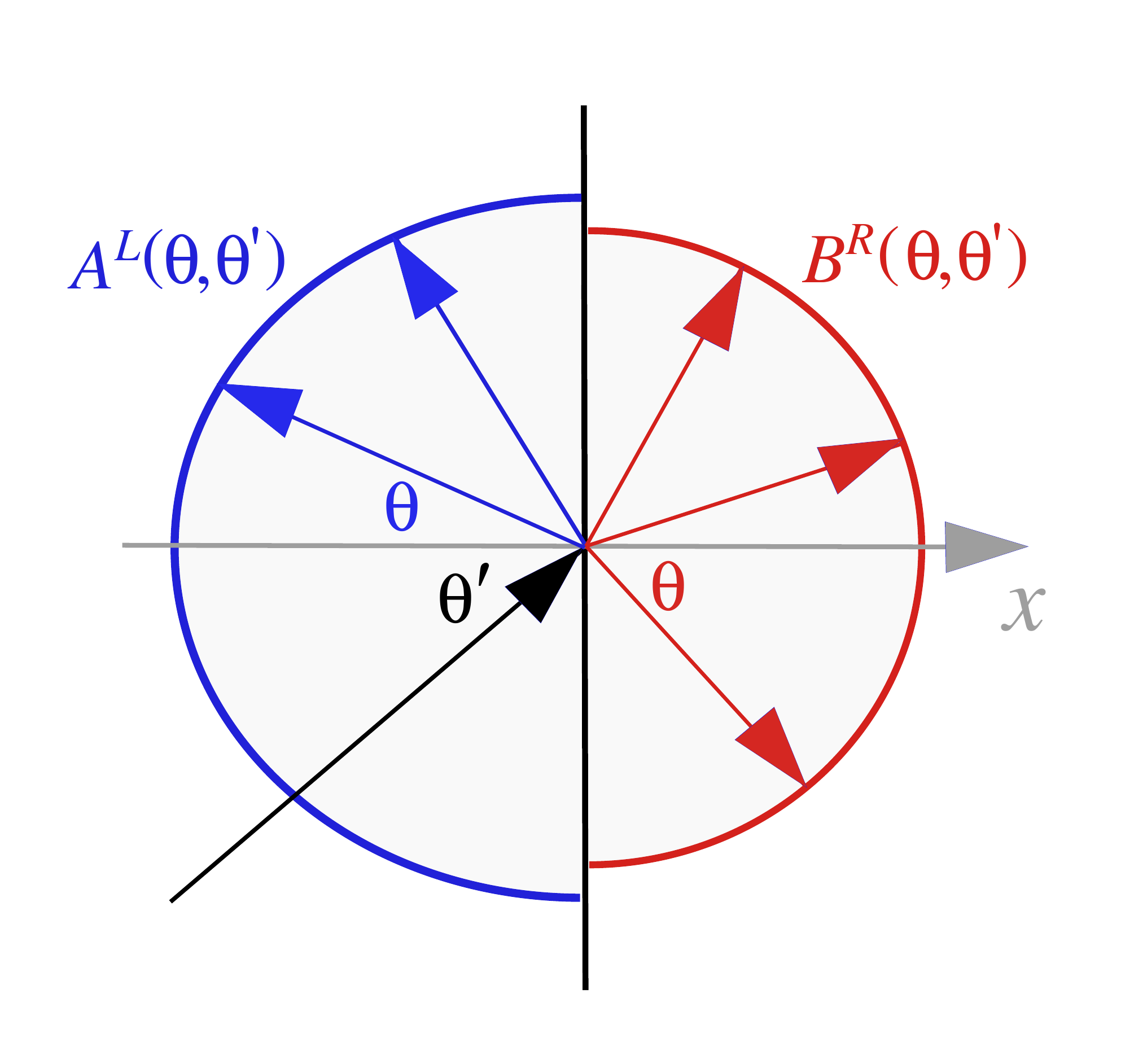} \caption{The interface is shown as a bald vertical line. The black arrow depicts
a wave with a unit amplitude that is incident on the interface from
the left. Colored arrows represent all the reflected and transmitted
waves. $A^{L}$-s are amplitudes of waves reflected on the left side,
and $B^{R}$-s are amplitudes of waves transmitted to the right. Those
amplitudes are functions of $\theta'$ -- the angle of incidence
and $\theta$ -- angles of departure, which are also shown. Angles
are calculated from the $x$ axis, which is perpendicular to the interface,
shown as a gray arrow. Colored circles illustrate our model: for given
$\theta'$ all amplitudes on one side have the same magnitude (Eq.
\ref{EqDMM}).}
\label{Fig3} 
\end{figure}

To calculate coefficients in formulae (\ref{EqMs}) we want to calculate
jumps of temperature and electrochemical potential at the interface
$\Delta T$ and $\Delta\zeta$ given heat flux and electric current
through the interface. There are contributions to these jumps directly
associated with the reflection of electrons at the interface $\Delta T^{B}$
and $\Delta\zeta^{B}$. Since the distribution function of electrons
in the vicinity of the interface is perturbed by the interface, there
arise additional effective contributions to jumps $\Delta T^{L,R}$
and $\Delta\zeta^{L,R}$. These effective contributions arise on both
sides of the interface and indexes $L,R$ denote the left or the right
side.

We consider an interface between two n-type semiconductors with a
simple zone structure. We assume the bottom of the conduction band
is higher for the semiconductor on the right. We call the energy difference
between those bottoms an energy barrier $V_{b}$, and we place the
origin of the energy axis at the bottom of the conduction band of
the left crystal. Thus, the dispersion relations for the left and
right semiconductors are 
\begin{align}
\varepsilon & =\frac{p^{L^{2}}}{2m^{L}}\nonumber \\
\varepsilon & =\frac{p^{R^{2}}}{2m^{R}}+V_{b}.\label{EqDispersion}
\end{align}
where $\varepsilon$ is the energy of an electron, $m^{L,R}$ and
$p^{L,R}$ are effective masses and momentums of electrons on the
left and the right sides, respectively.

To perform our calculations, we use the system of equations introduced
in Ref. \cite{KinDMM} and methods from there, adapted for electrons.
We want to note that the model described in \cite{KinDMM} is much
simpler and all the calculations there can be very easily verified
by pen and paper, while in this paper all integrals and systems of
equations have the very same physical meaning but can be solved only
numerically. Therefore, in order to understand further calculations,
it may be wise to read the manuscript \cite{KinDMM} first.

The central feature of the method is the introduction of matching
equations for the distribution functions at the interface. We briefly
explain the idea here. We seek the solution of the wave equation (Schrodinger
equation in the case of electrons) as a superposition of the incident
wave and a set of reflected and transmitted waves with amplitudes
$A$-s and $B$-s that we call reflection and transmission amplitudes
(Fig. \ref{Fig3}). We use squares of the amplitudes as the coefficients
for the matching equations of the distribution functions at the interface.

The concept of matching equations is only applicable if electrons
don't scatter inelastically in the barrier region. This gives another
condition for the applicability of the method: the width of the barrier
is less than the electron mean free pass.

The matching equations for the distribution functions of electrons
($n$) at the interface are 
\begin{align}
n^{L\leftarrow}(\theta)= & \int_{0}^{1}d\cos\theta'|A_{\theta\theta^{'}}^{L}|^{2}n^{L\rightarrow}(\theta^{'})+\nonumber \\
+\frac{p^{R}m^{R}}{p^{L}m^{L}} & \int_{0}^{1}d\cos\theta'|B_{\theta\theta^{'}}^{L}|^{2}n^{R\leftarrow}(\theta^{'})\nonumber \\
n^{R\rightarrow}(\theta)= & \int_{0}^{1}d\cos\theta'|A_{\theta\theta^{'}}^{R}|^{2}n^{R\leftarrow}(\theta^{'})|+\nonumber \\
+\frac{p^{L}m^{L}}{p^{R}m^{R}} & \int_{0}^{1}d\cos\theta'|B_{\theta\theta^{'}}^{R}|^{2}n^{L\rightarrow}(\theta^{'}),\label{EqMatching}
\end{align}
where arrows denote the direction of propagation, $\theta'$ -- the
angle of incidence, $\theta$ -- the angle of reflection or transmission,
angles are counted from the $x$ axis perpendicular to the interface.

The derivation of matching equations for electrons is very similar
to  the one for phonons, provided in \cite{Me}. This derivation however
is based on quantum mechanics rather than kinetic theory.
For the sake of not over-complicating the present paper with different
methods, this derivation will be published elsewhere. In this manuscript,
we  focus on kinetic theory only.

To describe the distribution function of electrons near the interface,
we use the conventional stationary Boltzmann equations in the relaxation
time approximation: 
\begin{equation}
\frac{p^{L,R}}{m^{L,R}}\cos\theta\frac{\partial n^{L,R}}{\partial x}+eE^{*}\frac{\partial n^{L,R}}{\partial p}=-\frac{\chi^{L,R}}{\tau^{L,R}}.\label{EqBoltzmann}
\end{equation}

Here $\tau^{L,R}$ are electron relaxation times on the left and right
sides of the interface. $n_{0}$ -- is an equilibrium distribution
function, which is the Fermi-Dirac distribution. $\chi^{L,R}=n^{L,R}-n_{0}^{L,R}$
are nonequilibrium parts of the distribution functions. $E^{*}$ is
the effective electric field $E^{*}=\nabla U^{*}=E+\nabla\mu/e=\nabla\zeta/e$.

In a previous work \cite{KinDMM} we introduce the Chapman-Enskog
conditions \cite{LdKin}. Since the total number of electrons is conserved,
there are two such conditions, instead of one, as in  the case of
 phonons. To define the electrochemical potential of a non-equilibrium
system, we introduce the condition that the number of particles in
a non-equilibrium system is equal to the number of particles in an
equilibrium system with the same electrochemical potential: 
\begin{equation}
\int\frac{d^{3}p}{(2\pi\hbar)^{3}}\chi^{L,R}=0.\label{EqEnskog1}
\end{equation}
The temperature of a non-equilibrium system is defined as the temperature
of an equilibrium system with the same energy: 
\begin{equation}
\int\frac{d^{3}p}{(2\pi\hbar)^{3}}(\varepsilon-\zeta)\chi^{L,R}=0.\label{EqEnskog2}
\end{equation}

We also need the conservation of flow equations. Again, for electrons,
not only the energy but also the number of particles, or equivalently,
the electric charge is conserved. We have 
\begin{align}
2e\int\frac{d^{3}p}{(2\pi\hbar)^{3}}v_{x}^{R}\chi^{R}=j\nonumber \\
2\int\frac{d^{3}p}{(2\pi\hbar)^{3}}v_{x}^{R}(\varepsilon-\zeta)\chi^{R}=q.\label{EqFluxes}
\end{align}
Here $v_{x}^{R}=p^{R}/m^{R}$ is the electron velocity in the right
semiconductor. In the second equation, we only account for  the heat
energy, not the total amount of energy so we write $\varepsilon-\zeta$,
where $\zeta$ is electrochemical potential \cite{Zaim}. Because
of fluxes conservation at the interface, we can only write these two
equations for the right side. Analogous equations for the left side
at the very interface are fulfilled automatically.

The equations (\ref{EqMatching} - \ref{EqFluxes}) form the general
system of equations that describes the transport through the interface
of two n-type semiconductors with a simple zone structure. We want
to specify the model by choosing the amplitudes for the matching equations
(\ref{EqMatching}).

To simplify the calculations, we  define the
electron analog of the Density Mismatch Model (DMM) to describe the
electron scattering at the interface. The DMM is not a very accurate
model for phonons, but it simplifies the calculations greatly and
usually can predict correctly an order of magnitude \cite{Stoner}.
We think that a modification of this model can be useful for electron
transport calculations, as well.

To find conditions for the transmission and reflection amplitudes,
which constitute an electron analog of the DMM, we assume uniform
scattering at the interface. We also assume, as in Ref. \cite{KinDMM},
that the fraction of the energy flux dissipated in a certain direction
does not depend on the angle of incidence. It yields 
\begin{align}
|A_{\theta\theta'}^{L,R}|^{2}=|A^{L,R}|^{2}\cos\theta'\nonumber \\
|B_{\theta\theta'}^{L,R}|^{2}=|B^{L,R}|^{2}\cos\theta',\label{EqDMM}
\end{align}
We write down the flow conservation equation for each mode (Exactly
one mode is presented in Fig. \ref{Fig3}). After some simplifications,
they take the form 
\begin{align}
\cos\theta'= & \int_{0}^{1}d\cos\theta\cos\theta|A_{\theta\theta'}^{L}|^{2}+\nonumber \\
+\frac{p^{R}m^{L}}{p^{L}m^{R}} & \int_{0}^{1}d\cos\theta\cos\theta|B_{\theta\theta'}^{R}|^{2}\nonumber \\
\cos\theta'= & \int_{0}^{1}d\cos\theta\cos\theta|A_{\theta\theta'}^{R}|^{2}+\nonumber \\
+\frac{p^{L}m^{R}}{p^{R}m^{L}} & \int_{0}^{1}d\cos\theta\cos\theta|B_{\theta\theta'}^{L}|^{2}.\label{EqUnitFlow}
\end{align}

We substitute the equilibrium distribution functions into equation
(\ref{EqMatching}), substitute the conditions (\ref{EqDMM}) into
equations (\ref{EqMatching}, \ref{EqUnitFlow}) and perform the angle
integration. Thus we find the set of equations to determine values
of reflection and transmission amplitudes. By solving it we find 
\begin{align}
|A^{L}|^{2} & =\frac{2p^{L^{2}}}{p^{L^{2}}+p^{R^{2}}}\nonumber \\
|A^{R}|^{2} & =\frac{2p^{R^{2}}}{p^{L^{2}}+p^{R^{2}}}\nonumber \\
|B^{L}|^{2} & =\frac{2m^{L}p^{L}p^{R}}{m^{R}(p^{L^{2}}+p^{R^{2}})}\nonumber \\
|B^{R}|^{2} & =\frac{2m^{R}p^{L}p^{R}}{m^{L}(p^{L^{2}}+p^{R^{2}})}.\label{EqSetAmplitudes}
\end{align}

Now we solve the system of equations (\ref{EqMatching} - \ref{EqFluxes})
with the dispersion relations (\ref{EqDispersion}) and the set of
amplitudes (\ref{EqDMM}, \ref{EqSetAmplitudes}).

Boltzmann equation (\ref{EqBoltzmann}) describes the evolution of
the electron distribution function near the interface. We divide the
solution to the Boltzmann equation into complementary and particular
parts, $\chi^{R}=\chi_{p}^{R}+\chi_{c}^{R}$. The complementary solution
is: 
\begin{equation}
\chi_{c}^{R}=\chi_{0}^{R}\exp{(-x/v_{x}^{R}\tau^{R})},\label{EqChiC}
\end{equation}
where $\chi_{0}^{R}=\chi^{R}(x=0)$. Similarly, for the left crystal
we have $\chi_{c}^{L}=\chi_{0}^{L}\exp{(x/v_{x}^{L}\tau^{L})}$. We
observe that for incident electrons the solution of this type increases
infinitely. The complementary part of the solution does not satisfy
the boundedness condition, which means that the distribution function
of incident electrons is determined only by a particular solution.

The particular solution at the interface is 
\begin{align}
\chi_{p}^{L,R}=- & \tau^{L,R}v_{x}^{L,R}\frac{\varepsilon-\zeta}{kT^{2}}\left(\frac{dT}{dx}\right)_{0}^{L,R}-\nonumber \\
- & \tau^{L,R}v_{x}^{L,R}\frac{1}{kT}\left(\frac{d\zeta}{dx}\right)_{0}^{L,R}.\label{EqChiP0}
\end{align}
At the proximity of interface $\left(\frac{dT}{dx}\right)^{L,R}$
and $\left(\frac{d\zeta}{dx}\right)^{L,R}$ -- are unknown functions
of the coordinate $x$, since, due to the perturbation of the
electron distribution functions by the interface, the temperature
and the electrochemical potential gradients near the interface differ
from the gradients in a homogeneous media. So we have six unknown
parameters that characterize the distribution function of electrons
at the interface These are the temperature and the electrochemical
potential jumps at the interface $\Delta T^{B},\Delta\zeta^{B}$,
and four gradients at the very interface on both sides of the interface.

We substitute the expression (\ref{EqChiP0}) into the matching equations
(\ref{EqMatching}) and obtain the distribution function for receding
electrons. Now we know the full distribution function of electrons
at the very interface, expressed with six unknown parameters. We substitute
it into equations (\ref{EqEnskog1}, \ref{EqEnskog2}, \ref{EqFluxes}).
Now we have a system of six equations for six variables.

Here we can see the necessity of two jumps at the interface, the temperature
jump $\Delta T$ and the electro-chemical potential jump $\Delta U^{*}$.
Each of these jumps is associated with three unknown parameters for
our system of six equations. If we do not introduce $\Delta U^{*}$,
we do not have enough unknown parameters for six equations. Since
the equations (\ref{EqEnskog1}, \ref{EqEnskog2}, \ref{EqFluxes}),
that distribution function at the interface should satisfy, are basically
the conservation of energy and the electric charge equations, we believe
that our prediction of the existence of electrochemical potential
jump at the interface is based on solid grounds.

We solve the equations (\ref{EqEnskog1}, \ref{EqEnskog2}, \ref{EqFluxes})
to  find the jumps and the gradients at the very interface. They are
expressed with two externally given parameters $q,j$ that are introduced
in the equations (\ref{EqFluxes}). Here we see why we used inverse
proportionality coefficients in equations (\ref{EqMs}). In our calculations
we naturally use the fluxes $q,j$ to compute the jumps $\Delta T$,
$\Delta U^{*}$, not the otherwise.

Now that we know the gradients at the interface, we can find the effective
jumps associated with them. Let us find the solution of the Boltzmann
equation for the  crystal on the right, since the solution for that
of the  left  is completely analogous. From here on, we omit the index
$R$.

We divide the gradients into two parts: 
\begin{align}
\left(\frac{dT}{dx}\right)=\left(\frac{dT}{dx}\right)_{p}+\left(\frac{dT}{dx}\right)_{\infty}\nonumber \\
\left(\frac{d\zeta}{dx}\right)=\left(\frac{d\zeta}{dx}\right)_{p}+\left(\frac{d\zeta}{dx}\right)_{\infty},\label{EqGrads}
\end{align}
where index $\infty$ indicates the gradient at an infinite distance
from the interface and $p$ stands for perturbed, which is the difference
between the gradient at a given point and infinity. Now for the particular
part of the solution for receding electrons, we write 
\begin{align}
\chi_{p}=- & \tau v_{x}\frac{\varepsilon-\zeta}{kT^{2}}\left[\left(\frac{dT}{dx}\right)_{p}+\left(\frac{dT}{dx}\right)_{\infty}\right]-\nonumber \\
- & \tau v_{x}\frac{1}{kT}\left[\left(\frac{d\zeta}{dx}\right)_{p}+\left(\frac{d\zeta}{dx}\right)_{\infty}\right].\label{EqChiP1}
\end{align}
We substitute this expression (\ref{EqChiP1}) and the expression
for the complementary part of the nonequilibrium function (\ref{EqChiC})
into expressions for the heat flux and electric current (\ref{EqFluxes}).
For the heat flux, we obtain 
\begin{align}
2\int\frac{d^{3}k}{(2\pi)^{3}}v_{x}(\varepsilon-\zeta)(\tau v_{x}\frac{\varepsilon-\zeta}{kT^{2}}\left[\left(\frac{dT}{dx}\right)_{p}+\left(\frac{dT}{dx}\right)_{\infty}\right]+\nonumber \\
+\tau v_{x}\frac{1}{kT}\left[\left(\frac{d\zeta}{dx}\right)_{p}+\left(\frac{d\zeta}{dx}\right)_{\infty}\right]+\chi_{c})=q.
\end{align}
Now we observe, that integral expression behind $\left(\frac{dT}{dx}\right)$
and $\left(\frac{d\zeta}{dx}\right)$ are coefficients $L_{TT}$ and
$L_{TE}$ (\ref{EqLs}) in relaxation time approximation. Since at
infinity, the heat flux is the same as in the homogeneous media 
\begin{equation}
q=L_{TT}\left(\frac{dT}{dx}\right)_{\infty}+L_{TE}\left(\frac{d\zeta}{dx}\right)_{\infty},\label{EqInfty}
\end{equation}
we can subtract this from both sides. Now we obtain 
\begin{align}
2\int\frac{d^{3}k}{(2\pi)^{3}}v_{x}(\varepsilon-\zeta)\left(\tau v_{x}\frac{d\chi_{p}}{dx}+\chi_{c}\right)=\nonumber \\
=L_{TT}\left(\frac{dT}{dx}\right)_{p}+L_{TE}\left(\frac{d\zeta}{dx}\right)_{p}.
\end{align}
We integrate  $x$ from zero to infinity. Since this integration of
the perturbed part of the gradients gives, by definition, effective
jumps, we obtain 
\begin{equation}
2\int_{0}^{\infty}dx\int\frac{d^{3}k}{(2\pi)^{3}}v_{x}(\varepsilon-\zeta)\chi_{c}=L_{TT}\Delta T^{R}+L_{TE}\Delta\zeta^{R}.\label{EqDeltaQ}
\end{equation}

We perform analogous manipulations with the expression for electric
current in Eq. (\ref{EqFluxes}) and obtain 
\begin{equation}
2e\int_{0}^{\infty}dx\int\frac{d^{3}k}{(2\pi)^{3}}v_{x}\chi_{c}=L_{ET}\Delta T^{R}+L_{EE}\Delta\zeta^{R}.\label{EqDeltaJ}
\end{equation}

We substitute the expression for the complementary part of the distribution
function (\ref{EqChiC}) into expressions (\ref{EqDeltaQ}, \ref{EqDeltaJ})
and perform the integration in the left part. Now, we have a system
of two equations for $\Delta T^{R},\Delta\zeta^{R}$, so we find them.

We can also find $\Delta T^{L},\Delta\zeta^{L}$ in the same manner,
but the calculation is a bit longer, since we should also account
for electrons below the energy barrier, and substitute their
distribution function into expressions for heat flux and electric
current.

Now we can sum and obtain the full temperature jump $\Delta T=\Delta T^{L}+\Delta T^{B}+\Delta T^{R}$
and the full electrochemical potential jump $\Delta\zeta=\Delta\zeta^{L}+\Delta\zeta^{B}+\Delta\zeta^{R}$.
We recall that $\Delta U^{*}=\Delta\zeta/e$, so we have found the
effective potential jump. The proportionality coefficients between
jumps $\Delta T,\Delta U^{*}$ and fluxes $q,j$ are coefficients
from the formulae (\ref{EqMs}).

We have counted all the coefficients describing linear transport phenomena
at the interface between n-type semiconductors. Let us note the nonlinear
one. The second equation from (\ref{EqFluxes}) is only equal on both
sides in the linear approximation. Since a jump of $\zeta$ at the
interface occurs, heat fluxes differ on both sides of the interface.
The difference between fluxes 
\begin{equation}
q^{L}-q^{R}=2\int\frac{d^{3}k}{(2\pi)^{3}}v_{x}^{R}\chi^{R}\,\Delta\zeta
\end{equation}
is released at the interface resulting in itsheating up. Since $\chi$
has components, that are proportional to $\Delta T$ and $\Delta\zeta$,
we have two components of heat release. Oneof them, proportional to
$\Delta\zeta^{2}$ is the interfacial analog of the Joule effect,
and the other one, proportional to $\Delta T\Delta\zeta$, is the
analog of the Thomson effect.

\section{Results and discussion}

We perform calculations for the interface between two samples of Ga$_{x}$In$_{1-x}$As
with different values of $x$: $x^{L}$ and $x^{R}$. The energy barrier
$V_{b}$ is given by the difference between affinities plus the difference
between Fermi levels. The affinity is given by the formula $4.9-0.83x$
eV \cite{GaInAs1}. Materials with different values of $x$ also have
different effective masses, whose values are given by $0.023+0.037x+0.003x^{2}\ m_{0}$
\cite{GaInAs2}.

\begin{figure}[t]
\centering \includegraphics[width=0.49\textwidth]{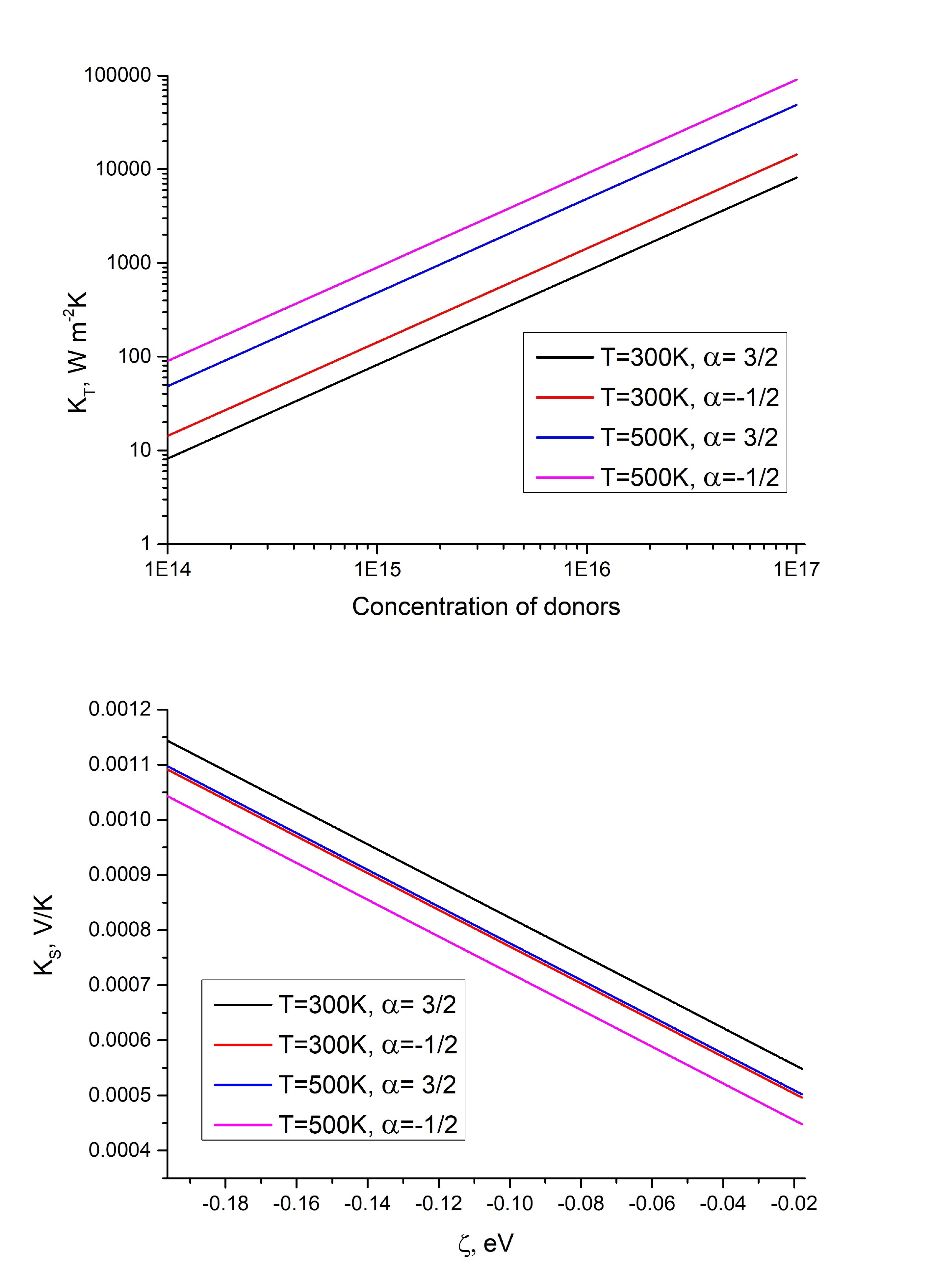} \caption{Dependencies of kinetic coefficients $K_{T},K_{S}$ (\ref{EqKs})
on the concentration of donors, and electrochemical potential $\zeta$,
respectively, at different temperatures and for different power laws
of a relaxation time $\tau(\varepsilon)$: $\tau\sim\varepsilon^{\alpha}$.
It is clearly observed that $K_{T}$ depends linearly on concentration,
while $K_{S}$ depends linearly on $\zeta$. Power law of $\tau(\varepsilon)$
switches from $\alpha=-1/2$ to $\alpha=3/2$ with the growth of concentration,
so experimental data should be approximated correctly by the $-1/2$
law at low concentration, and by $3/2$ law at high concentration.}
\label{GraphsND} 
\end{figure}

The Fermi level is defined by the concentration of donors
$n_{D}$. The compact formula for the Fermi level is different for
different temperatures,  donor concentrations, and  energies of donor
ionization $\varepsilon_{D}$. We will assume here that donors are
shallow such as Sn, Ge, Si. Typical values of $\varepsilon_{D}$ for
shallow donors is about 5 meV \cite{GaInAs3}. Since we are interested
in applications, we are mostly concerned with the temperatures about
room temperature and higher. These temperature values satisfy the
condition $kT\approx\varepsilon_{D}$ or even $kT>\varepsilon_{D}$.
For such conditions the following equation \cite{Anselm} holds: 
\begin{equation}
\zeta=kT\ln\frac{4\pi^{2}\hbar^{3}n_{D}}{(2\pi mkT)^{3/2}}.\label{EqZeta}
\end{equation}
At these  conditions all donors are ionized, which means that the
concentration of electrons is equal to the concentration of donors.
It also implies that scattering on donors is the scattering on charged
impurities, not on the neutral ones.

The difference in the Fermi levels on both sides causes band bending
and the occurrence of the space charge region at the interface. The
region is small if the Fermi levels difference is small. If the space
charge region is small compared to the electron free pass, the theory
is applicable. The occurrence of space charge should in principle
enhance scattering on the interface, but in our model, we assume scattering
is already maximal (Eq. \ref{EqDMM}).
Therefore,  the calculation remains unchanged. If the space charge
region is thick enough, the theory should be  modified. That is the
case for the well-known p-n junction theory, where transport in the
space charge region is considered to be diffusive \cite{Anselm}.

\begin{figure}[t]
\centering \includegraphics[width=0.49\textwidth]{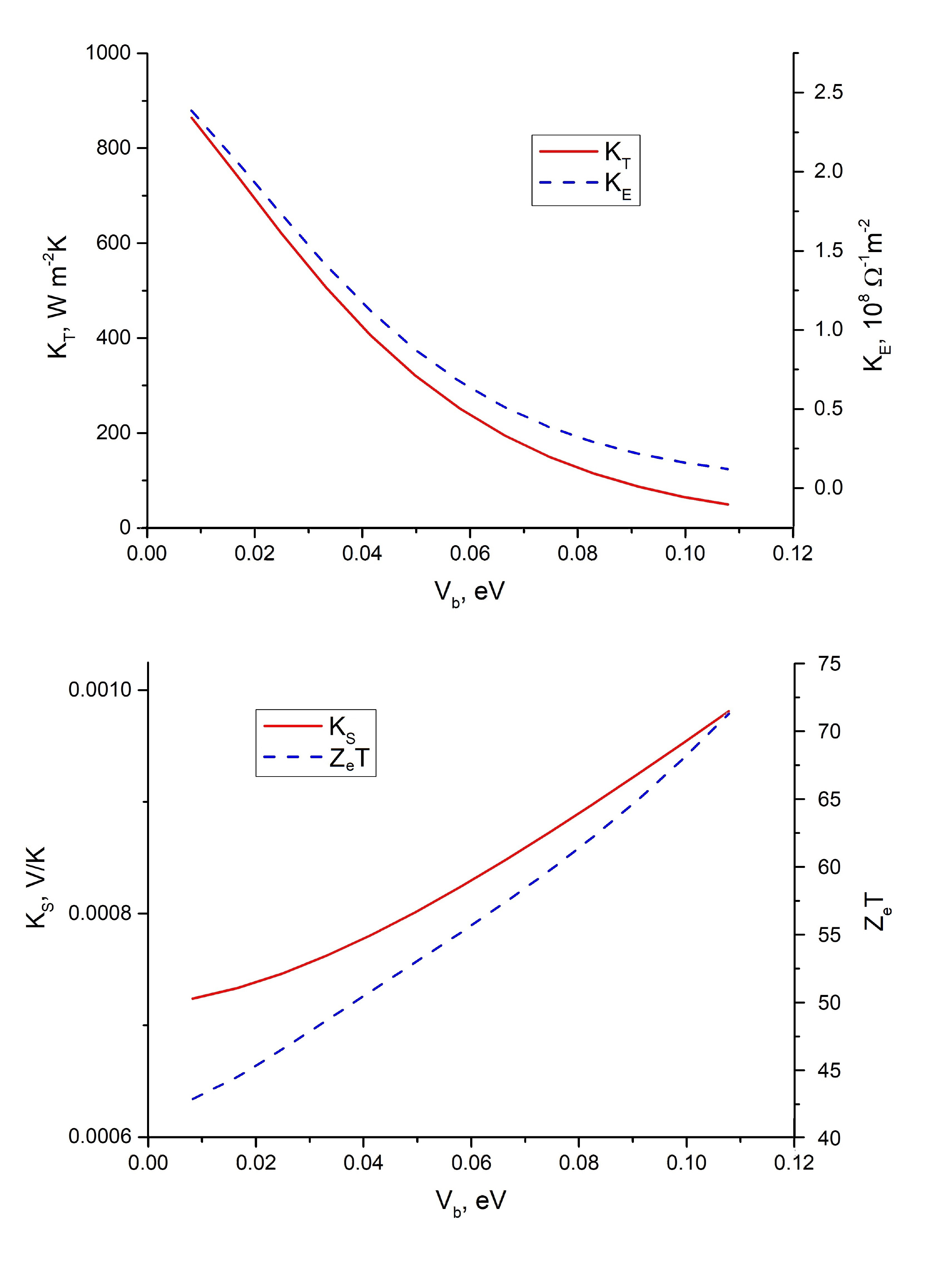} \caption{On the upper graph dependencies of kinetic coefficients $K_{T},K_{E}$
(\ref{EqKs}) on the energy barrier height are shown. On the lower
graph, another kinetic coefficient $K_{S}$ together with electronic
thermoelectric figures of merit $Z_{e}T$ versus energy barrier are
depicted. We can clearly see that both $K_{T}$ and $K_{E}$ drop
similarly as the energy barrier value grows, while $K_{S}$ increases.
This leads to the overall growth of $Z_{e}T$. It is important to
note here that $Z_{e}T$ (Eqs. \ref{EqZeT1}, \ref{EqZeT2})  only
considers the electronic transport properties and does not take into
account phonon heat transfer. }
\label{GraphsVb} 
\end{figure}

Finally, before presenting the calculation results, we  address the
relaxation times that we use for our calculations. The absolute value
of the relaxation time does not affect the values of the computed
temperatures and the electrochemical potential jumps. Higher relaxation
time leads to a higher gradient at the interface, but also to a faster
relaxation of the disturbance. Two effects cancel out. This is, indeed
the case,  for a much simpler model that has been treated in Ref.
\cite{KinDMM}. Independence of $\Delta T^{L,R}$ and $\Delta\zeta^{L,R}$
on the intrinsic properties of the materials is one reason to treat
them as contributions to interfacial jumps.

However, the dependence of electron relaxation time on energy does
affect the result, since it affects the form of the distribution function
at the interface. Relaxation time dependence on energy $\tau(\varepsilon)$
is related to   the main scattering source. If scattering is mainly
produced by   phonons, the dependence is $\tau\sim\varepsilon^{-1/2}$.
When scattering is dominated by  charged impurities the dependence
is $\tau\sim\varepsilon^{3/2}$ \cite{Anselm}. For low concentration
of impurities, phonons are the main source of scattering, while for
a high concentration of impurities, impurity scattering dominates.
The relaxation time $\tau(\varepsilon)$ must be changed accordingly.

First, we present dependencies on the concentration of donors, since
those are most clearly understandable. For simplicity, here we assume
that the concentration of donors in both materials is  such, that
Fermi levels in both materials are the same. We vary concentration
in the left semiconductor $n_{D}^{L}$ in the range from $10^{14}$
cm$^{-3}$ to $10^{17}$ cm$^{-3}$, and adjust the concentration
in the right conductor, accordingly. Lower values of concentration
do not even produce n-type semiconductors. With higher values, the
Fermi level comes too close to the conductivity band, and we have
to include Fermi statistics together with possible many-body effects.

All theoretic coefficients, M-s (\ref{EqMs}) depend linearly on the
concentration of electrons, which is equal to the concentration of
donors in the concerned conditions. Thismeans that both, $K_{T}$
and $K_{E}$ depend linearly on concentration. However, $K_{S}$ is
a quotient of $M_{ET}$ and $M_{TT}$, and their dependence on concentration
cancels out. $K_{S}$ is proportional to the mean heat energy carried
by each electron, which is approximately $kT-\zeta$. As it is shown
by the formula (\ref{EqZeta}), $\zeta$ depends logarithmically on
concentration. Both of these dependencies are presented in Figure
\ref{GraphsND}. The graph of $K_{E}$ is not presented since it is
just the same as $K_{T}$.

We calculate $K_{T},K_{E}$, and $K_{S}$, but to characterize the
efficiency of the interface for application as a thermoelectric generator,
we also  calculate the thermoelectric figure of merit. For a thermoelectric
material, we have $ZT=S^{2}\sigma T/\kappa$, where $S$ is Seebeck
coefficient, $\sigma$ and $\kappa$ are electrical and thermal conductivities,
respectively. $ZT$ is the most important parameter of the thermoelectric
material \cite{ThermEl6}. In analogy with the homogeneous case, for
the interface, we can write 
\begin{equation}
Z_{e}T=\frac{K_{S}^{2}K_{E}}{K_{T}}T.\label{EqZeT1}
\end{equation}
We also can express it in terms of theoreticaly computable parameters
(\ref{EqMs}): 
\begin{equation}
Z_{e}T=\frac{M_{ET}M_{TE}}{M_{TT}M_{EE}-M_{ET}M_{TE}}.\label{EqZeT2}
\end{equation}
 It is important to note here, that in this paper, we only calculate
the electronic part of heat conduction, and with it, we calculate
$Z_{e}T$, which characterizes the electronic subsystem. As we will
see in a few paragraphs, heat conductance by phonons through the interfaces
under consideration, is a few orders of magnitude higher, than that
of electrons. So, actual  $ZT$ values for given interfaces should
be quite small. We only present our calculations of $Z_{e}T$ here
to demonstrate its dependence on different parameters of the interface.

\begin{figure}[t]
\centering \includegraphics[width=0.49\textwidth]{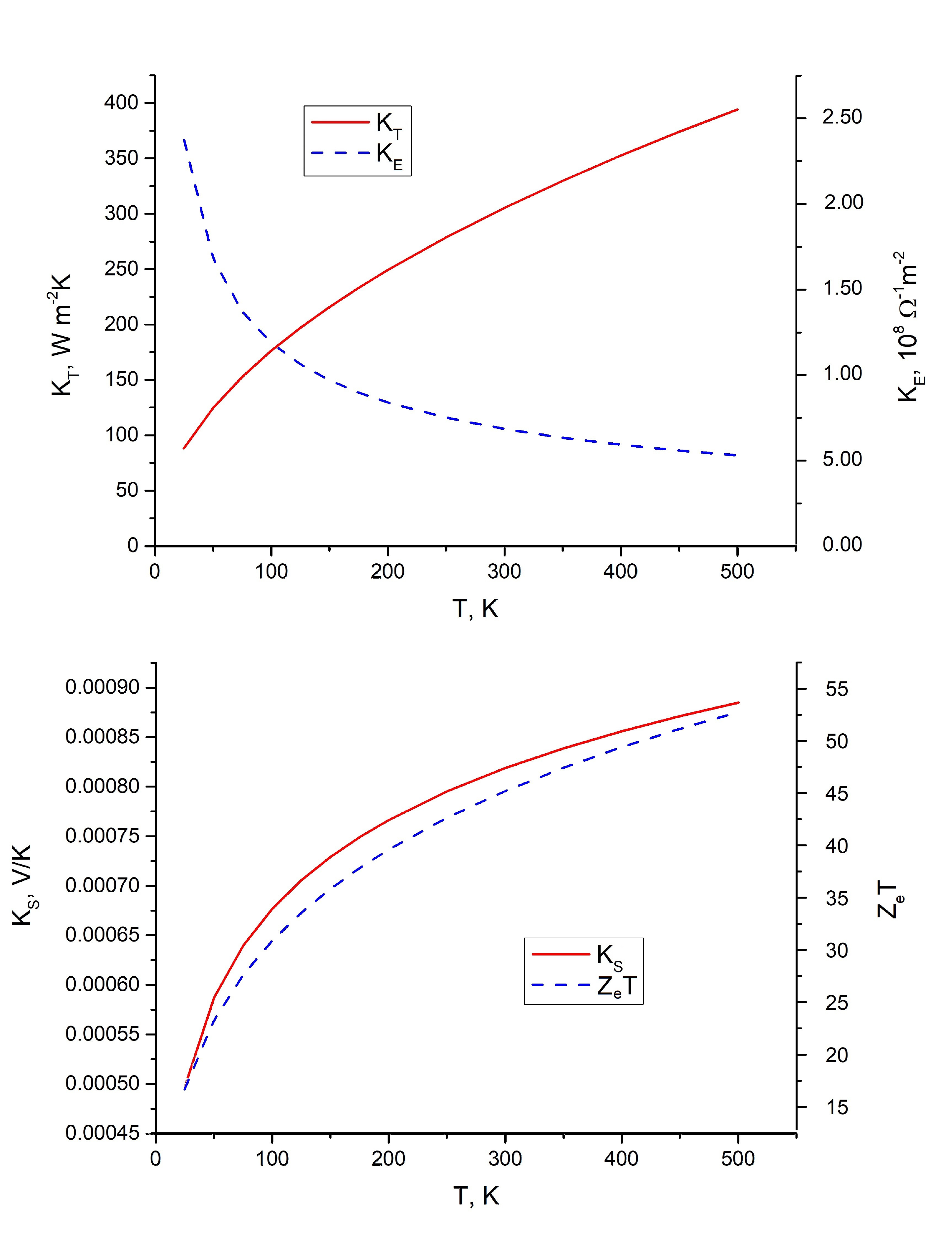} \caption{On the upper graph dependencies of kinetic coefficients $K_{T},K_{E}$
(\ref{EqKs}) on the temperature are shown.  The kinetic coefficient
$K_{S}$ and $Z_{e}T$ versus temperature are shown on the lower graph.
It can be observed that, while $K_{T}$ grows with temperature, $K_{E}$
decreases. Also, both $K_{S}$ and $Z_{e}T$ increase with temperature.
It is important to note that $Z_{e}T$ (Eqs. \ref{EqZeT1}, \ref{EqZeT2})
here only considers the electronic transport properties and does not
take into account phonon heat transfer.}
\label{GraphsT} 
\end{figure}

In Figure \ref{GraphsVb} the coefficients $K_{T},K_{E},K_{S}$, and
$ZT$ are shown as functions of the height of the barrier. Graphs
are plotted for $T=300$ K, $n_{D}^{L}=10^{15}$, $x^{L}=0.47$. $x^{R}$
is varied from 0.48 up to 0.6 thus creating different heights of energy
barrier. Again we adjust the concentration of donors in the right
conductor so that Fermi levels in both materials are the same. We
can see, that by variating the height of the barrier, we can get different
values of the parameters $K_{T}$ and $K_{E}$. We think, that for
small barriers we underestimate the values of $K_{T}$ and $K_{E}$,
since for very alike materials' reflection at the interface is very
small, and inthe limit of similar materials it should vanish. This
leads to infinite values of all the $M$-s from formulae (\ref{EqMs}),
which turn out in  infinite values of $K_{T}$ and $K_{E}$ at such
a limit. Finite values of $K_{T}$ and $K_{E}$ at $V_{b}=0$ are
artifacts of the chosen DMM-like model. Away from zero, the DMM-like
model works fine and correctly predicts the reduction  of the values
of $K_{T}$ and $K_{E},$ with a growth of $V_{b}$. This is caused
by the increased reflection upon increasing $V_{b}$. Also, both $K_{T}$
and $K_{E}$ have a very similar dependence on the barrier height.
On the other hand, from the lower graph of Figure \ref{GraphsVb},
we note that $K_{S}$ grows with the growth of the barrier. And because
$K_{S}$ grows and $K_{T}$, $K_{E}$ diminish, the electronic part
of $ZT$ grows quite fast with the growth of $V_{b}$. This means
that higher barriers are better suited to use for thermoelectric purposes.

Finally,  present temperature dependencies. In this case, we can not
just assume the equality of Fermi levels on both sides of the interface,
since Fermi level and temperature are related by equation (\ref{EqZeta}).
If Fermi levels are equal at one temperature, they will not be equal
at all the other temperatures. The second thing to note, is that,
as the formula (\ref{EqZeta}) shows, the height of the potential
barrier $V_{b}$ vanishes at the zero temperature limit. Indeed, at
zero temperature the Fermi level tends to the bottom of the conductivity
band, and after $\zeta$  equalizes on both sides, the bottoms of
conductivity bands would be on the same level. There would still remain
scattering, because of the spacial charge and the mismatch of effective
masses on both sides. However, as we have shown on the  upper
graph of Fig. \ref{GraphsVb}, all conductivities increase significantly
with the decrease of $V_{b}$. Additionally,  changes in temperature
have another effect, besides changing $V_{b}$. Heat conductivity
also increases with the mean heat energy of particles, which is about
$kT-\zeta$, and grows linearly with temperature. For the interfacial
heat conductance $K_{T}$, this effect turns out to be stronger, than
the drop caused by $V_{b}$ growth. Therefore, $K_{E}$ lowers with
temperature while $K_{T}$ grows with temperature. This result can
be seen in Figure \ref{GraphsT}. It is also observed that  $K_{S}$
and $ZT$ grow with temperature, which we could not predict with only
qualitative reasoning.

We calculated these coefficients for a range of different values of
donor concentration, heights of the energy barriers between semiconductors,
and different temperatures. For $K_{T}$ we have found typical values
to be $10^{2}-10^{3}$Wm$^{-2}$K$^{-1}$. These are very small numbers
compared to typical values of Kapitza conductances known for phonon
transport, which are $10^{7}-10^{8}$Wm$^{-2}$K$^{-1}$ \cite{Stoner}.

Let us show that this discrepancy of 5 orders of magnitude is due
to very different concentration of carriers in the cases of electron
and phonon interfacial transport: 
\begin{equation}
\frac{K_{T-e}}{K_{T-ph}}\approx\frac{n_{D}}{N},\label{EqEsteem}
\end{equation}
where $K_{T-e},K_{T-ph}$ are electron and phonon Kapitza thermal
conductances, $N$ is the concentration of atoms in the crystal lattice.

Indeed at a temperature about the Debye temperature, the amount of
phonons is about the number of atoms in the lattice. The typical concentration
of atoms in the crystal lattice is about $10^{22}$ cm$^{-3}$. For
semiconductors in the conditions that we study, the amount of electrons
is equal to the concentration of donors, for example, $10^{16}$ cm$^{-3}$
for $K_{T}$ to be equal to $10^{3}$Wm$^{-2}$K$^{-1}$ at room temperature
(Fig.\ref{Fig3}). The quotient of these concentrations gives $10^{6}$,
which is an order of magnitude different from the quotient of the
boundary thermal conductivities. Given that it is a very rough estimate,
we conclude that Eq.\ref{EqEsteem} is a valid approximation and our
calculated values of $K_{T}$ are reasonable.

Kinetic coefficients of semiconductors are that high, despite low
concentrations of electrons, because of the small effective masses
of electrons and large relaxation times. To say it short, because
of high electron mobility. But the values of Kapitza parameters do
not depend on absolute values of effective masses and relaxation times.
That is why a low concentration of electrons results in a low value
of the Kapitza parameters. That shows that heat transport through
the interface between semiconductors is mostly due to phonons.

For $K_{E}$, typical values are about $10^{8}$ $\Omega^{-1}$m$^{-2}$.
We have no reference to compare it to, but we can estimate the correctness
by the Wiedemann-Franz law. Linear law of dependence on temperature
does not work in the considered case, because the characteristic amount of heat
energy carried by one electron is not $kT$, but $kT-\zeta$. However,
we can substitute $kT-\zeta$ into the Wiedemann-Franz law and use
it as an order of magnitude estimate for $K_{E}$. 
\begin{equation}
\frac{K_{T}}{K_{E}}\approx\frac{k(kT-\zeta)}{e^{2}}.\label{EqWiedFr1}
\end{equation}
Substituting $K_{T}$ and $K_{E}$, calculated for different sets
of parameters into this formula, we find that indeed the expression
on the left and the expression on the right are always of the same
order of magnitude. More precisely we have found 
\begin{equation}
\frac{K_{T}}{K_{E}}=L\frac{k(kT-\zeta)}{e^{2}},\label{EqWiedFr1}
\end{equation}
where $L$ is a dimensionless coefficient. Values of $L$ for different
parameters vary in the range from 1.5 to 3.

For $K_{S}$ they are $10^{-4}-10^{-3}$ V/K. That is a very high
value, since the highest known values are about $2\times10^{-3}$,
 for Pb$_{15}$Ge$_{37}$Se$_{58}$
\cite{RecordTE}. Moreover, while for calculations of $K_{T}$ and
$K_{E}$, we only hope to find the correct order of magnitude, the
calculation of $K_{S}$ is probably more precise. That is because
the simple model of uniform scattering at the interface inevitably
produces incorrect factors for the theoretic coefficients M-s (\ref{EqMs}).
But these factors should be about equal for all the M-s. Since $K_{S}$
is a quotient of two M-s, these factors cancel out and the result
can be quite accurate.

\section{Conclusions}

The main result of the present manuscript is the prediction that for
electronic subsystems, separated by an interface, in a non-equilibrium
state, two jumps on the interface should arise: the temperature jump
and the electrochemical potential jump. The existence of a temperature
jump is well known, at least for phonon transport. In this paper,
we introduce the electrochemical potential jump. We provide two different arguments for the
existence of such a jump. The first one is based on non-equilibrium thermodynamics and the other one is based on kinetics.

We have also presented a description of linear kinetic processes at
the interface between n-type semiconductors, in terms of three kinetic
coefficients. These are interfacial analogs of electric and heat conductances
$K_{E}$ and $K_{T}$ and the interfacial analog of the Seebeck coefficient
$K_{S}$. These coefficients are important for the description of
nanostructured materials, where the lengths between interfaces are
small compared to the so-called Kapitza length, i.e.  the quotient
between the kinetic coefficient of the interface to the related kinetic
coefficient of the homogeneous media.

 In addition, we developed  a method for the calculation of interfacial
kinetic coefficients. To the authors best knowledge all three coefficients
$K_{E}$, $K_{T}$ and $K_{S}$ that characterize electronic transport
through an interface are calculated here for the first time.

We presented the results of our calculations in the form of dependencies
of kinetic coefficients on donor concentration, heights of the energy
barriers between semiconductors, and temperature. We have found that
the interfacial analog of the Seebeck coefficient $K_{S}$, for some
range of parameters, has a high value of about $10^{-3}$ V/K,, only twice smaller than the highest ever observed value.

There is a deep reason, why $K_{S}$ can have a very high value for different types
of semiconductor interfaces. For n-type semiconductors, $K_{S}$
is proportional to the absolute value of quasi-Fermi level $\zeta$,
counted from the bottom of the conduction band (see Fig. \ref{GraphsND}).
That  means that we can increase the value of $\zeta$ to increase
the value of the coefficient $K_{S}$. But if the Fermi level is placed
deep inside the energy gap, there exists a large number of holes.
These produce thermoelectric currents in the opposite direction to
that of the electrons  canceling out the overall thermoelectric effect.
However, it can be chosen  two semiconductors such that an interface
between them would have a low barrier for electrons and a very high
barrier for holes, thus blocking the hole transport. This will allow
the superstructure consisting of such semiconductors to have a Fermi
level very deep inside the band gap, while having almost no hole current,
which in turn will make a material with a very high value of Seebeck
coefficient.

For materials with high acousticmismatch  a  lower phonon heat conductance
is obtained, thus increasing the thermoelectric figure of merit. All
these properties  give a good opportunity to produce a superlattice
with record high thermoelectric parameters.

\acknowledgments{ The author is grateful for A.\,Ya. Vul and C.
Pastorino for their attention to the presented investigation.}

\end{document}